\documentclass[journal]{IEEEtran}

\usepackage{color}
\usepackage{balance}
\usepackage{graphicx}
\usepackage{enumerate}
\usepackage{subfigure}
\usepackage{algorithm} 
\usepackage{algorithmic}
\usepackage{booktabs}
\usepackage{cite}
\usepackage{amssymb}
\usepackage{amsmath} 
\usepackage{stfloats}
\usepackage{url}

\begin{document}

\title{Deep Learning-Aided OFDM-Based Generalized Optical Quadrature Spatial Modulation}

\author{
Chen Chen, \IEEEmembership{Member, IEEE,} Lin Zeng, Xin Zhong, Shu Fu, Min Liu, and Pengfei Du

\thanks{This work was supported in part by the National Natural Science Foundation of China under Grant 61901065 and in part by the Project funded by China Postdoctoral Science Foundation under Grant 2021M693744.}
\thanks{C. Chen, L. Zeng, X. Zhong, S. Fu, and M. Liu are with the School of Microelectronics and Communication Engineering, Chongqing University, Chongqing 400044, China (e-mail: \{c.chen, 202012021008, 201912131098, shufu, liumin\}@cqu.edu.cn).}
\thanks{P. Du is with the A*STAR's Singapore Institute of Manufacturing Technology (SIMTech), Singapore 138634 (e-mail: Du\_Pengfei@simtech.a-star.edu.sg).}
}

\maketitle

\begin{abstract}
In this paper, we propose an orthogonal frequency division multiplexing (OFDM)-based generalized optical quadrature spatial modulation (GOQSM) technique for multiple-input multiple-output optical wireless communication (MIMO-OWC) systems. Considering the error propagation and noise amplification effects when applying maximum likelihood and maximum ratio combining (ML-MRC)-based detection, we further propose a deep neural network (DNN)-aided detection for OFDM-based GOQSM systems. The proposed DNN-aided detection scheme performs the GOQSM detection in a joint manner, which can efficiently eliminate the adverse effects of both error propagation and noise amplification. The obtained simulation results successfully verify the superiority of the deep learning-aided OFDM-based GOQSM technique for high-speed MIMO-OWC systems.

\end{abstract}


\section{Introduction}

\IEEEPARstart{W}{ith} the explosive increasing of mobile data traffic in recent years, traditional radio frequency communication technologies such as WiFi might not be able to support the heavy mobile data traffic in the near future. Owing to its abundant spectrum resource, optical wireless communication (OWC) exploring visible light, infrared and ultra-violet spectrum has been emerging as a promising candidate to meet the requirement of the ever-increasing mobile data traffic \cite{cogalan2017would}. Although the spectrum resource of OWC systems is abundant, the achievable data rate of practical OWC systems is largely limited by the small modulation bandwidth of the optical components, especially the optical transmitters. For the OWC systems using commercial off-the-shelf light-emitting diodes (LEDs), the -3dB bandwidth is usually only a few MHz, which greatly limits the achievable capacity of the system \cite{rajagopal2012ieee}.

To boost the capacity of bandlimited OWC systems, various spectral efficiency enhancing techniques have already been proposed so far. Among them, multiple-input multiple-output (MIMO) transmission and orthogonal frequency division multiplexing (OFDM) modulation are considered as two most promising techniques \cite{fath2013performance,mesleh2011performance}. As a digitized MIMO scheme, optical spatial modulation (OSM) has drawn great attention in OWC systems lately. In OSM, only a single optical transmitter is selected to transmit constellation symbols and the spectral efficiency of OSM systems is contributed by both the constellation symbols and the spatial index symbols \cite{mesleh2011optical}. As a result, OSM systems have the advantages of negligible inter-channel interference, high power efficiency and low transceiver complexity \cite{soltani2019bidirectional}. Nevertheless, the achievable spectral efficiency of OSM systems is mainly limited by the number of the optical transmitters and it is challenging for practical OSM systems to achieve high spectral efficiencies. In order to increase the spectral efficiency contributed by the spatial index symbols and hence further enhance the spectral efficiency of OSM systems, generalized OSM (GOSM) has been proposed where multiple optical transmitters are selected to simultaneously transmit the same constellation symbol \cite{wang2020constellation}. Furthermore, the combination of both OFDM and various OSM techniques can be a more efficient way to enhance the spectral efficiency of OWC systems. Specifically, OFDM-based OSM has been reported in \cite{Yesilkaya2019ofdm,chen2021enhanced}, while OFDM-based GOSM has been proposed in \cite{chen2020ofdm}.

In this paper, we for the first time propose a novel OFDM-based generalized optical quadrature spatial modulation (GOQSM) technique for further spectral efficiency enhancement of bandlimited MIMO-OWC systems. The proposed GOQSM is inspired by the concept of quadrature spatial modulation (QSM) \cite{mesleh2014quadrature}, where spatial mapping can be performed in both the in-phase and quadrature components and hence a doubled number of spatial bits can be transmitted. As a result, the proposed GOQSM can be considered as a generalized version of QSM in OWC systems, which can provide a much higher spectral efficiency than that of GOSM. Moreover, we further propose two detection schemes for OFDM-based GOQSM systems, including maximum likelihood and maximum ratio combining (ML-MRC)-based detection and deep neural network (DNN)-aided detection. Simulation results show that the proposed OFDM-based GOQSM technique significantly outperforms OFDM-based GOSM when applying the deep learning-aided detection scheme.

\section{System Model}

In this paper, we consider a general MIMO-OWC system equipped with $N_t$ LEDs and $N_r$ photo-diodes (PDs). Letting $\bf{s}$ = $[s_{1}, s_{2}, \cdots, s_{N_t}]^T$ be the transmitted signal vector, $\bf{H}$ be the $N_r \times N_t$ MIMO channel matrix and $\bf{n}$ = $[n_{1}, n_{2}, \cdots, n_{N_r}]^T$ be the additive noise vector, the received signal vector $\bf{y}$ = $[y_{1}, y_{2}, \cdots, y_{N_r}]^T$ can be expressed by 
	\begin{equation}
	\setlength{\abovedisplayskip}{12pt}
	\setlength{\belowdisplayskip}{12pt}
	\bf y = H s + n,
	\label{eq:y}
	\end{equation}  
where the channel matrix of the $N_r \times N_t$ MIMO-OWC system can be given by 
	\begin{equation}
	\setlength{\abovedisplayskip}{12pt}
	\setlength{\belowdisplayskip}{12pt}
	\bf H = \left[\begin{array}{ccc}
	h_{1 1} & \cdots & h_{1 N_t}\\
	\vdots & \ddots & \vdots\\
	h_{N_r 1} & \cdots & h_{N_r N_t}
	\end{array}
	\right],
	\label{eq:H}
	\end{equation}
where $h_{r t}$ ($r = 1, 2, \cdots, N_r; t = 1, 2, \cdots, N_t$) denotes the direct current (DC) channel gain between the $t$-th LED and the $r$-th PD. In this work, we assume that each LED follows the Lambertian radiation pattern and only the line-of-sight (LOS) transmission is considered \cite{komine2004fundamental}. Hence, $h_{r t}$ can be calculated as follows:
	\begin{equation}
	\setlength{\abovedisplayskip}{12pt}
	\setlength{\belowdisplayskip}{12pt}
	h_{rt} = \frac{(l+1) \rho A}{2\pi d_{rt}^2} \mathrm{cos}^l(\varphi_{rt}) T_s(\theta_{rt}) g(\theta_{rt})   \mathrm{cos}(\theta_{rt}),
	\label{eq:h}	
	\end{equation}  	
where $l = -\mathrm{ln}2/\mathrm{ln}(\mathrm{cos}(\Psi))$ represents the Lambertian emission order and $\Psi$ is the semi-angle at half power of the LED; $\rho$ and A denote the responsivity and the active area of the PD, respectively; the distance, the angle of emission and the angle of incidence between the $t$-th LED and the $r$-th PD are expressed by $d_{rt}$, $ \varphi_{rt}$ and $\theta_{rt}$, respectively; $T_s (\theta_{rt})$ is the gain of optical filter; $g(\theta_{rt}) = \frac{n^2}{\mathrm{sin}^2\Phi}$ is the gain of optical lens, where $n$ and $\Phi$ denote the refractive index and the half-angle field-of view (FOV) of the optical lens, respectively.  

In MIMO-OWC systems, the additive noise is generally the sum of both shot and thermal noises and it can be reasonably modeled as a real-valued zero-mean additive white Gaussian noise (AWGN) with power $P_{n} = N_0 B$, where $N_{0}$ and ${B}$ represent the noise power spectral density (PSD) and the signal bandwidth, respectively.    

	\begin{figure*}[!t]
	\centering
	{\includegraphics[width=0.99\textwidth]{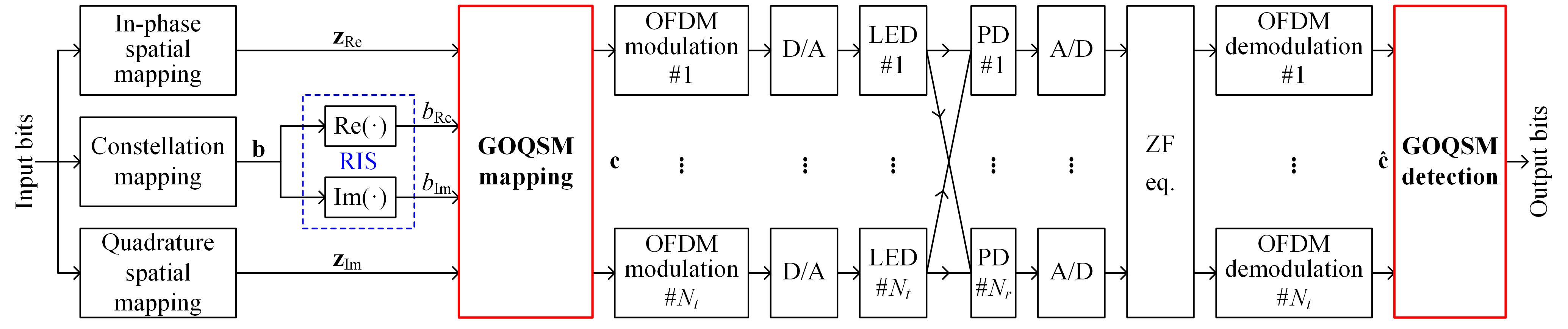}}
	\caption{Schematic diagram of the proposed OFDM-based GOQSM system.} 
	\label{fig:1}
	\end{figure*}
     
\section{OFDM-Based GOQSM}

In this section, we first describe the fundamental principle of OFDM-based GOQSM, and then we propose two detection schemes for OFDM-based GOQSM systems, including maximum likelihood and maximum ratio combining (ML-MRC)-based detection and DNN-aided detection.  

\subsection{Principle of OFDM-Based GOQSM}

Fig. \ref{fig:1} shows the schematic diagram of the proposed OFDM-based GOQSM system with $N_t$ LEDs and $N_r$ PDs. As we can see, the incoming bit stream is divided into three parts: one part is mapped into a constellation symbol vector $\textbf{b}$, while the other two parts are mapped into an in-phase spatial index vector $\textbf{z}_{\textrm{Re}}$ and a quadrature spatial index vector $\textbf{z}_{\textrm{Im}}$, respectively. The complex-valued element $b$ of the constellation symbol vector $\textbf{b}$ can be represented as $b = b_{\textrm{Re}} + j \times b_{\textrm{Im}}$, where $b_{\textrm{Re}}$ and $b_{\textrm{Im}}$ denote the real and imaginary parts of $b$, respectively. Before performing GOQSM mapping, the real and imaginary parts of $b$ are separated through real-and-imaginary separation (RIS). In OFDM-based GOQSM systems, GOQSM mapping is performed at each subcarrier slot in the frequency domain with respect to both the real and imaginary components. More specifically, for an OFDM-based GOQSM system with totally $N_t$ LEDs and $N$ ($N \le N_t$) activated LEDs, $N$ out of $N_t$ real parts of the subcarrier-modulating symbols corresponding to $N_t$ LEDs at each subcarrier slot are selected to transmit $b_{\textrm{Re}}$ while the remaining $N_t - N$ real parts are set to zero. At the same time, $N$ out of $N_t$ imaginary parts of the subcarrier-modulating symbols corresponding to $N_t$ LEDs at each subcarrier slot are selected to transmit $b_{\textrm{Im}}$ while the remaining $N_t - N$ imaginary parts are set to zero. Assuming $\textbf{c}_{\textrm{Re}}$ and $\textbf{c}_{\textrm{Im}}$ are the resultant real and imaginary vectors, the finally transmitted symbol vector can be obtained as  $\textbf{c} = \textbf{c}_{\textrm{Re}} + j \textbf{c}_{\textrm{Im}}$. After that, parallel OFDM modulation and digital-to-analog (D/A) conversion are performed and the resultant signals are used to drive $N_t$ LEDs, respectively.     

At the receiver side, the optical signal is first detected by $N_r$ PDs and then the obtained electrical signals are digitized via analog-to-digital (A/D) conversion. Subsequently, zero-forcing (ZF) equalization and parallel OFDM demodulation are performed to generate the estimate of the transmitted symbol vector $\hat{\textbf{c}}$. To further estimate the constellation symbol vector and spatial index vector, two detection schemes including ML-MRC detection and DNN-aided detection are proposed, which are introduced in detail in the following subsection. 

For OFDM-based GOQSM systems using $M$-ary quadrature amplitude modulation (QAM) constellation with totally $N_t$ LEDs where $N$ out of $N_t$ LEDs are activated to transmit data, the achievable spectral efficiency is given by
	\begin{equation}
	\setlength{\abovedisplayskip}{12pt}
	\setlength{\belowdisplayskip}{12pt}
	R_{\textrm{GOQSM}} \approx \frac{1}{2} \log _{2}{(M)} +  \lfloor \log_ {2}{C(N_t , N)} \rfloor,
	\label{eq:R_QGSM}
	\end{equation}
where $\lfloor \cdot \rfloor$ represents the floor operation and $C(\cdot,\cdot)$ denotes the binomial coefficient. In (\ref{eq:R_QGSM}), the first term on the right-hand side is contributed by the constellation symbols where the scaling factor $\frac{1}{2}$ is due to the imposed Hermitian symmetry constraint for real-valued time-domain signal generation \cite{chen2020ofdm}, while the second term is contributed by both the in-phase and quadrature spatial index symbols. For the purpose of comparison, the achievable spectral efficiency of OFDM-based GOSM systems with exactly the same setup is given as follows \cite{chen2020ofdm}:
	\begin{equation}
	\setlength{\abovedisplayskip}{12pt}
	\setlength{\belowdisplayskip}{12pt}
	R_{\textrm{GOSM}} \approx \frac{1}{2} \log _{2}{(M)} + \frac{1}{2} \lfloor \log_ {2}{C(N_t , N)} \rfloor.
	\label{eq:R_GSM}
	\end{equation} 
It can be clearly found that, by replacing OSM with QOSM, the spectral efficiency contributed by the constellation symbols remains the same while the spectral efficiency contributed by the spatial index symbols is doubled.

\subsection{Two Detection Schemes for OFDM-Based GOQSM}

	\begin{figure}[!t]
	\centering
	\subfigure[ML-MRC detection]
	{\includegraphics[width=.99\columnwidth]{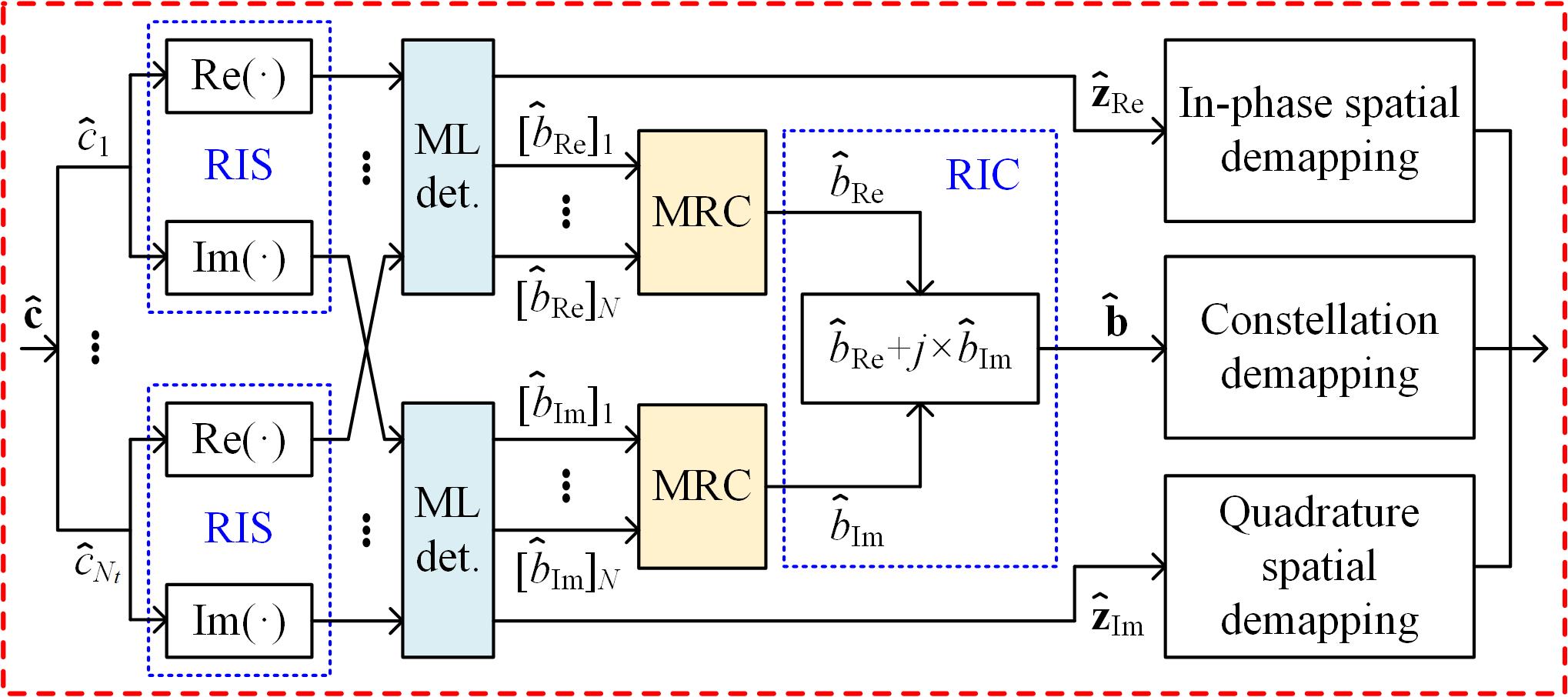}} 
	\quad
	\subfigure[DNN-aided detection]
	{\includegraphics[width=.99\columnwidth]{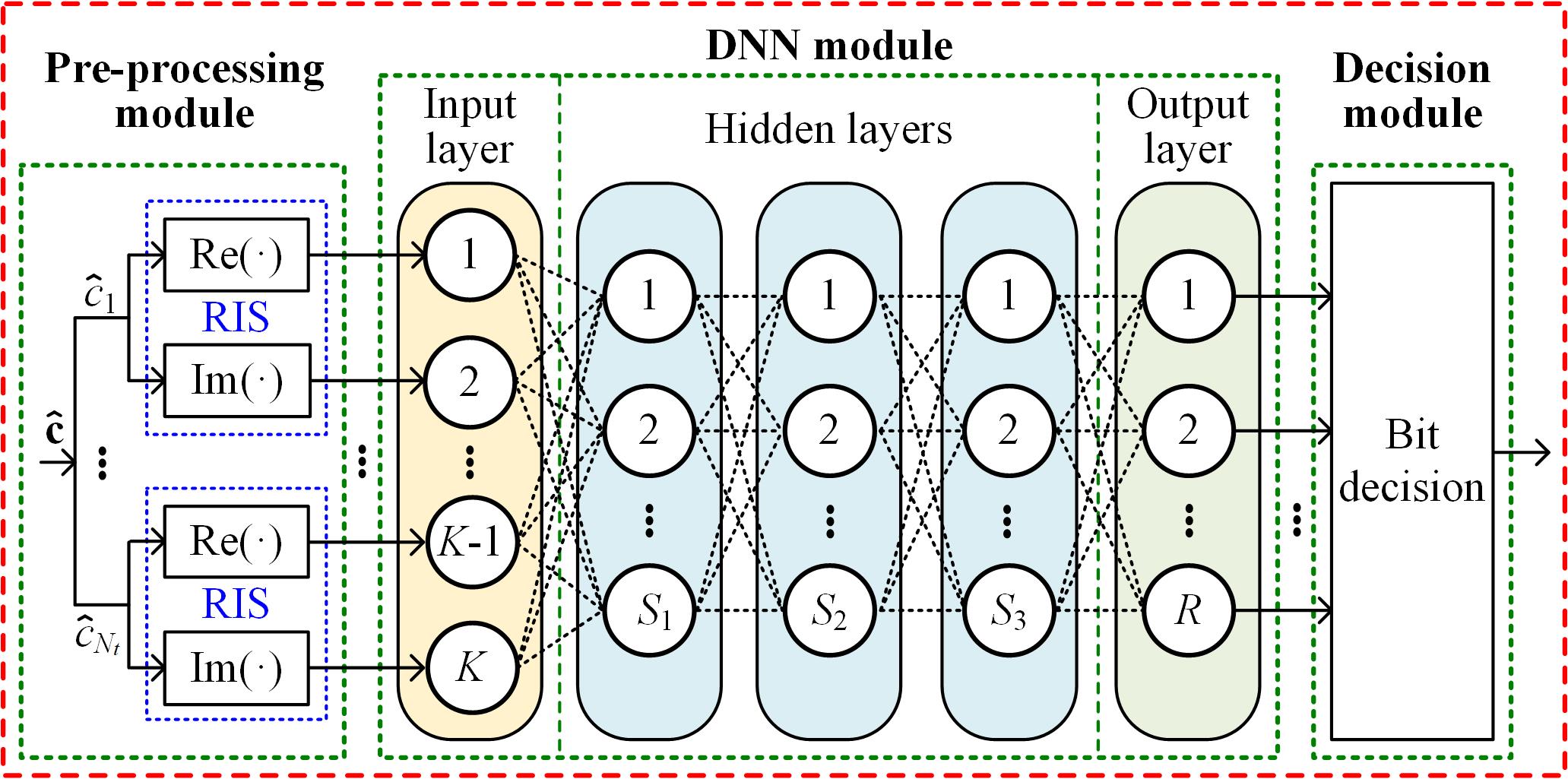}}
	\caption{Schematic diagrams of (a) ML-MRC detection and (b) DNN-aided detection.}
	\label{fig:2}
	\end{figure}	

As shown in Fig. \ref{fig:1}, GOQSM detection is performed to generate the final output bits after parallel OFDM demodulation. In the following, two GOQSM detection schemes are proposed including ML-MRC detection and DNN-aided detection.

\subsubsection{ML-MRC Detection}
The schematic diagram of the ML-MRC detection is illustrated in Fig. \ref{fig:2}(a), where the estimate of the transmitted symbol vector $\hat{\textbf{c}}$ = $[\hat{c}_1, \hat{c}_2, \cdots, \hat{c}_{N_t}]^T$ is treated as the input. Firstly, the real and imaginary parts of $\hat{c}_t$ ($t = 1, 2, \cdots, N_t$) are separated via RIS. Then, ML detection is performed with respect to both the real and imaginary parts, so as to obtain the corresponding estimates of the real components $[\hat{b}_{\textrm{Re}}]_m$, the imaginary components $[\hat{b}_{\textrm{Im}}]_m$, and the spatial index vectors $\hat{\textbf{z}}_{\textrm{Re}}$ and $\hat{\textbf{z}}_{\textrm{Im}}$. The detailed procedures to perform ML detection can be found in \cite{chen2020ofdm}, which is omitted here for brevity. After that, the estimates of $b_{\textrm{Re}}$ and $b_{\textrm{Im}}$, i.e, $\hat{b}_{\textrm{Re}}$ and $\hat{b}_{\textrm{Im}}$, can be obtained via MRC, and the estimate of the constellation symbol vector $\hat{\textbf{b}}$ can be further obtained by combining $\hat{b}_{\textrm{Re}}$ and $\hat{b}_{\textrm{Im}}$ via real-and-imaginary combination (RIC). Subsequently, in-phase and quadrature spatial demapping and constellation demapping are executed to recover the transmitted bit stream. 

As discussed above, the ML-MRC detection is a two-step detection scheme, and the errors occur during the first-step ML detection, if any, will directly propagate to the second-step MRC, resulting in error propagation. In the proposed OFDM-based GOQSM system, error propagation reduces the diversity gain that can be achieved by MRC and hence degrades the overall bit error rate (BER) performance of the system. Moreover, since the input $\hat{\textbf{c}}$ is generated after ZF equalization and parallel OFDM demodulation, the system also suffers from the adverse effect of noise amplification \cite{chen2020user}.
             
\subsubsection{DNN-Aided Detection}
To address the error propagation and noise amplification effects in the proposed OFDM-based GOQSM system using the ML-MRC detection, we propose a DNN-aided detection scheme for performance improvement of the OFDM-based GOQSM system in the following. Fig. \ref{fig:2}(b) illustrates the schematic diagram of the proposed DNN-aided detection scheme, which mainly consists of a pre-processing module, a feed-forward DNN module and a decision module. The pre-processing module is used to separate the real and imaginary parts of $\hat{c}_t$ via RIS and hence generate the input vector of the DNN module. For the OFDM-based GOQSM system with $N_t$ LEDs, the size of the input vector of the DNN module is given by $K = 2 N_t$. The feed-forward DNN module contains one input layer, three hidden layers and one output layer. The input layer is composed of $K$ neurons, which is corresponding to the size of the input vector. The numbers of the neurons of the three hidden layers are denoted by $S_1$ , $S_2$ and $S_3$, respectively. The three fully connected hidden layers are used to learn the statistical characteristics from the signals, where the rectified linear unit (ReLU) function, i.e., ${f_{\rm ReLU}(\alpha)}$ $=$ ${{\rm max}(0,\alpha)}$, is used as the activation function. The output layer is a fully connected layer with $R$ neurons, which utilizes the Sigmoid function, i.e., ${f_{\textrm{Sigmoid}}(\alpha)} = {1/(1+{\textrm{exp}}^{-\alpha})}$, as the activation function to map the output into the interval (0,1). Specifically, the value of $R$ is determined by the number of bits that can be transmitted by the input vector. Thus, we have $R = 2 R_{\textrm{GOQSM}}$ for OFDM-based GOQSM system. As a result, the input and output relationship of the feed-forward DNN module with totally five layers can be expressed by
   \begin{equation}
	\setlength{\abovedisplayskip}{12pt}
	\setlength{\belowdisplayskip}{12pt}
	{\bf{y}}_i = {\sigma}_i ({\bf{W}}_{i-1}{\bf{y}}_{i-1} + {\bf{b}}_{i-1}), ~~i = 2 , \cdots, 5,
	\label{eq:relationship}
	\end{equation}                 
where ${\sigma}_i (\cdot)$ is the activation function, and ${\bf{W}}_{i}$ and ${\bf{b}}_{i}$ denote the weight matrix and the bias vector, respectively.

Moreover, the decision module is utilized to generate the final binary bits. Letting ${\bf{y}}_5$ = $[y_1, y_2, \cdots, y_R]^T$ be the input of the decision module, the estimated bit information $\hat{g}_q$ ($q = 1, 2, \cdots, R$) can be expressed as follows:  
   \begin{equation}
	\setlength{\abovedisplayskip}{12pt}
	\setlength{\belowdisplayskip}{12pt}
	\hat{g}_q = 
	\begin{cases}
	0, & y_q < 0.5,\\
	1, & y_q \geq 0.5.  
	\end{cases} 
	\label{eq:estimate}
	\end{equation}     
Finally, the mean-squared error (MSE) loss function is applied to measure the difference between the transmitted bit vector ${\bf g}$ and the estimated bit vector ${\bf \hat{g}}$, which is given by 
	\begin{equation}
	\setlength{\abovedisplayskip}{12pt}
	\setlength{\belowdisplayskip}{12pt}
	e_{\textrm{MSE}} = \frac{1}{R}{\parallel {{\bf \hat{g}} {\bf -} {\bf g}} \parallel}^2.
	\label{eq:difference}
	\end{equation} 

\begin{table}[!t]
\centering
\caption{Simulation Parameters}\label{tab:Parameters}
\begin{tabular}{cc}
\toprule
Parameter & Value\\
\midrule
Room dimension & 5 m $\times$ 5 m $\times$ 3 m \\
LED spacing & 2.5 m \\
PD spacing & 10 cm \\
Semi-angle at half power of LED & $60^{\circ}$ \\
Gain of optical filter & 0.9 \\
Refractive index of optical lens & 1.5 \\
Half-angle FOV of optical lens & $72^{\circ}$ \\
Responsivity of the PD & 1 A/W \\
Height of receiving plane & 0.85 m \\
Active area of PD & 1 ${\textrm{cm}}^{2}$ \\
Noise PSD & $10^{-22} \textrm{A}^{2}$/Hz \\
Modulation bandwidth & 20 MHz \\
Number of LEDs & 4 \\
Number of activated LEDs & 2 \\
Number of PDs & 4 \\
Receiver location & (2 m, 2 m, 0.85 m) \\
\bottomrule
\end{tabular}
\end{table}

\begin{table}[!t]
\centering
\caption{Parameters of the DNN-Aided Detector}\label{tab:DNN}
\begin{tabular}{cc}
\toprule
Parameter & Value\\
\midrule
Number of input nodes & $K$=8 \\
Number of hidden layers & 3 \\
Hidden layer activation & ReLU \\
Output layer activation & Sigmoid \\
Loss function & MSE \\
Optimizer & Adam \\
Learning rate & 0.001 \\
Number of training set & 1524000 \\
Number of validation set & 635000 \\
Number of neurons@3 bits/s/Hz & $S_{1}$=30, $S_{2}$=36, $S_{3}$=16, $R$=6 \\
Number of neurons@4 bits/s/Hz & $S_{1}$=30, $S_{2}$=36, $S_{3}$=17, $R$=8 \\
Number of neurons@5 bits/s/Hz  & $S_{1}$=30, $S_{2}$=36, $S_{3}$=18, $R$=10 \\
\bottomrule
\end{tabular}
\end{table}
	
\section{Simulation Results}
In our simulations, we consider a 4$\times$4 MIMO-VLC system in a typical room with a dimension of 5 m $\times$ 5 m $\times$ 3 m. The four LEDs are mounted at the center of the ceiling and the height of the receiving plane is 0.85 m. Moreover, we compare the performance of OFDM-based GOQSM with OFDM-based GOSM using ML-MRC detection and DNN-aided detection. For both GOSM and GOQSM, two out of totally four LEDs are selected to transmit signals, i.e., $N_t = 4$ and $N = 2$. To achieve a target spectral efficiency under the above simulation setup, the required QAM constellations for GOQSM and GOSM can be obtained from (4) and (5), respectively. The other simulation parameters can be found in Table \ref{tab:Parameters}. Furthermore, the proposed DNN-aided detection is implemented on the PyTorch platform. In order to accelerate the training of the neural network, the mini-batch method is adopted and each mini-batch contains 100 symbol vectors. The training set and the validation set include $1.524 \times 10^{6}$ and $6.35 \times 10^{5}$ randomly generated symbol vectors, respectively. The key parameters of the DNN-aided detector are summarized in Table \ref{tab:DNN}.

	\begin{figure}[!t]
	\centering
	{\includegraphics[width=0.95\columnwidth]{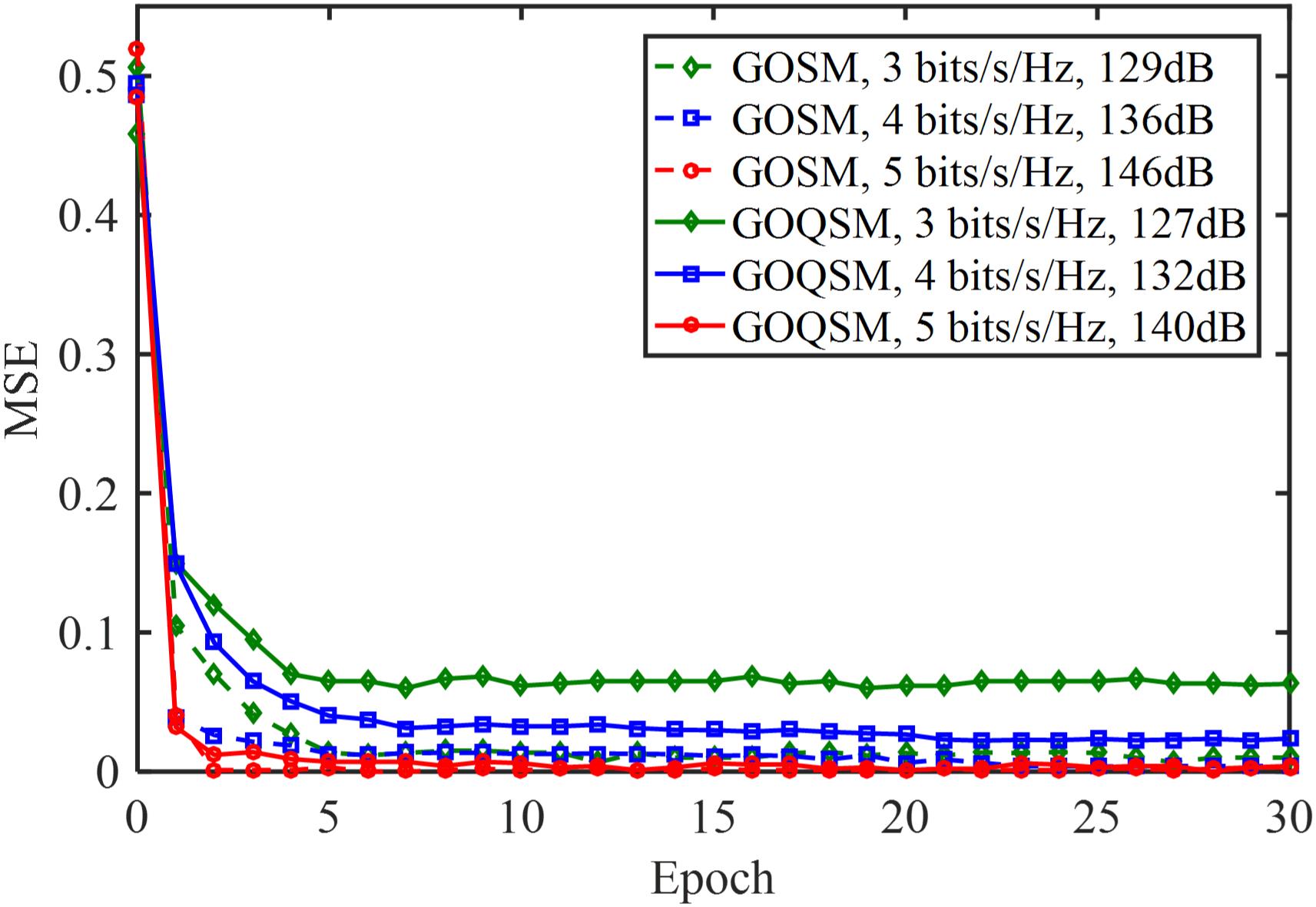}}
	\caption{MSE loss vs. epoch for the DNN-aided detector.} 
	\label{fig:MSE}
	\end{figure}

	\begin{figure}[!t]
	\centering
	{\includegraphics[width=0.95\columnwidth]{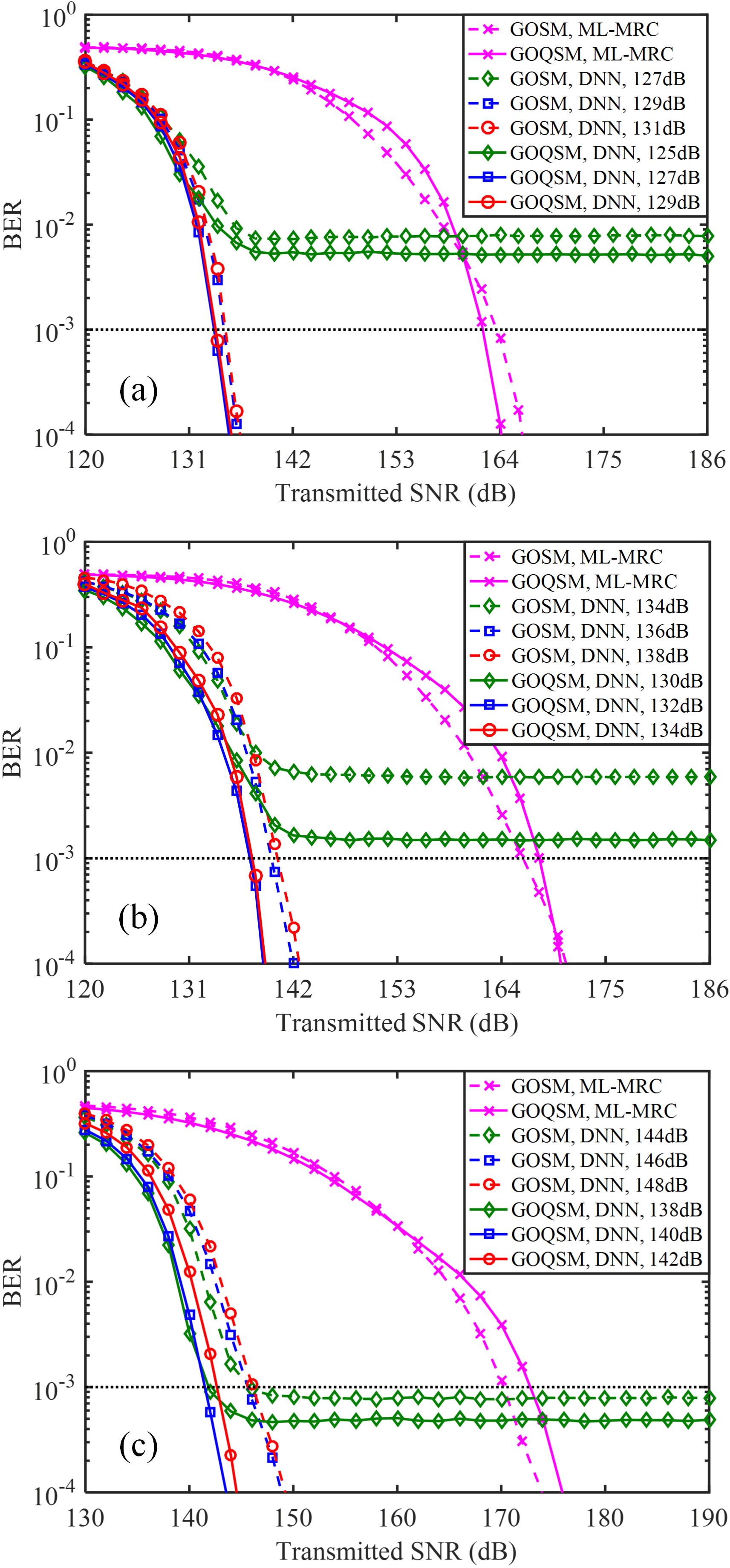}} 
	\caption{BER vs. transmitted SNR for OFDM-based GOSM and GOQSM with a spectral efficiency of (a) 3 bits/s/Hz, (b) 4 bits/s/Hz and (c) 5 bits/s/Hz.}
	\label{fig:3}
	\end{figure}

\subsection{MSE Loss}

We first analyze the MSE loss of the proposed DNN-aided detector and the MSE performance for GOSM and GOQSM is depicted in Fig. \ref{fig:MSE}. As we can observe, the MSE decreases rapidly with the increased number of epochs for both GOSM and GOQSM with different spectral efficiencies and different training signal-to-noise ratios (SNRs). Moreover, for a higher spectral efficiency, a larger training SNR is adopted for both GOSM and GOQSM. It can be clearly seen that the proposed DNN-aided detector achieves satisfactory performance within 10-epoch training for both GOSM and GOQSM with different spectral efficiencies and different training SNRs. Hence, the proposed DNN-aided detector can be implemented very quickly in practical systems.

\subsection{BER Performance}

In the next, we evaluate and compare the BER performance of OFDM-based GOSM and GOQSM systems with different spectral efficiencies. Figs. \ref{fig:3}(a)-(c) show the BER versus transmitted SNR with spectral efficiencies of 3, 4 and 5 bits/s/Hz, respectively. For a spectral efficiency of 3 bits/s/Hz, as shown in Fig. \ref{fig:3}(a), GOSM performs better than GOQSM in the small SNR region when using the ML-MRC detection and GOQSM outperforms GOSM when the SNR is larger than 160 dB. More specifically, the required SNRs for GOSM and GOQSM to achieve BER = $10^{-3}$ are 163.7 and 162.2 dB, respectively, indicating an SNR gain of 1.5 dB by applying GOQSM in comparison to GOSM. Moreover, when applying the proposed DNN-aided detection, there exists an optimal training SNR for both GOSM and GOQSM. This is because it is easier for the DNN to learn the statistics of the noise with a small training SNR and meanwhile the statistics of the data symbols can be more accurately learned by the DNN with a large training SNR. Therefore, the optimal training SNR can be seen as the trade-off for the DNN to learning the statistics of both the noise and the data symbols. It can be observed from Fig. \ref{fig:3}(a) that the optimal training SNRs for GOSM and GOQSM are 129 and 127 dB, respectively. With the optimal training SNR, GOQSM using the DNN-aided detection requires an SNR of 133.6 dB to reach BER = $10^{-3}$. Compared with GOQSM using the ML-MRC detection, a substantial SNR reduction of 28.6 dB is obtained, which is mainly due to the ability of the DNN-aided detector to eliminate noise amplification. Furthermore, GOQSM outperforms GOSM across the whole SNR region and it is mainly due to the elimination of error propagation by the DNN-aided detector. When applying the DNN-aided detection with an optimal training SNR, the SNR gain achieved by GOQSM in comparison to GOSM at BER = $10^{-3}$ is 1.1 dB. For higher spectral efficiencies of 4 and 5 bits/s/Hz, as shown in Figs. \ref{fig:3}(b) and (c), GOQSM performs worse than GOSM to achieve BER = $10^{-3}$ when using the ML-MRC detection. However, GOQSM performs much better than GOSM to reach BER = $10^{-3}$ when applying the DNN-aided detection. Specifically, the SNR gains are increased from 1.1 dB to 2.3 and 4.1 dB, when the spectral efficiency is increased from 3 bits/s/Hz to 4 and 5 bits/s/Hz, respectively.

\section{Conclusion}

In this paper, we have proposed an OFDM-based GOQSM scheme for MIMO-OWC systems. By applying QSM instead of SM, the achievable spectral efficiency of the system can be efficiently improved. In order to address the adverse effects of error propagation and noise amplification in the OFDM-based GOQSM system using the two-step ML-MRC detection, we have further proposed a DNN-aided detection scheme. Our simulation results show that the DNN-aided detection substantially outperforms the ML-MRC detection for both GOSM and GOQSM. Furthermore, a significant 4.1-dB SNR gain can be obtained by GOQSM in comparison to GOSM by using the DNN-aided detection. Therefore, OFDM-based GOQSM with deep learning-aided detection can be a promising candidate for high-speed MIMO-OWC systems.  

\bibliographystyle{IEEEtran}
\bibliography{IEEEabrv, mybib}

\begin{thebibliography}{10}
\providecommand{\url}[1]{#1}
\csname url@samestyle\endcsname
\providecommand{\newblock}{\relax}
\providecommand{\bibinfo}[2]{#2}
\providecommand{\BIBentrySTDinterwordspacing}{\spaceskip=0pt\relax}
\providecommand{\BIBentryALTinterwordstretchfactor}{4}
\providecommand{\BIBentryALTinterwordspacing}{\spaceskip=\fontdimen2\font plus
\BIBentryALTinterwordstretchfactor\fontdimen3\font minus
  \fontdimen4\font\relax}
\providecommand{\BIBforeignlanguage}[2]{{%
\expandafter\ifx\csname l@#1\endcsname\relax
\typeout{** WARNING: IEEEtran.bst: No hyphenation pattern has been}%
\typeout{** loaded for the language `#1'. Using the pattern for}%
\typeout{** the default language instead.}%
\else
\language=\csname l@#1\endcsname
\fi
#2}}
\providecommand{\BIBdecl}{\relax}
\BIBdecl

\bibitem{cogalan2017would}
T.~Cogalan and H.~Haas, ``\protect{Why would 5G need optical wireless
  communications?}'' in \emph{Proc. IEEE Ann. Int. Symp. Pers., Indoor Mobile
  Radio Commun. (PIMRC)}, Oct. 2017, pp. 1--6.

\bibitem{rajagopal2012ieee}
S.~Rajagopal, R.~D. Roberts, and S.-K. Lim, ``{IEEE 802.15. 7 visible light
  communication: modulation schemes and dimming support},'' \emph{IEEE Commun.
  Mag.}, vol.~50, no.~3, Mar. 2012.

\bibitem{fath2013performance}
T.~Fath and H.~Haas, ``{Performance comparison of MIMO techniques for optical
  wireless communications in indoor environments},'' \emph{IEEE Trans.
  Commun.}, vol.~61, no.~2, pp. 733--742, Feb. 2013.

\bibitem{mesleh2011performance}
R.~Mesleh, H.~Elgala, and H.~Haas, ``{On the performance of different OFDM
  based optical wireless communication systems},'' \emph{J. Opt. Commun.
  Netw.}, vol.~3, no.~8, pp. 620--628, Aug. 2011.

\bibitem{mesleh2011optical}
{R. Mesleh, H. Elgala, and H. Haas}, ``Optical spatial modulation,'' \emph{J.
  Opt. Commun. Netw.}, vol.~3, no.~3, pp. 234--244, Mar. 2011.

\bibitem{soltani2019bidirectional}
M.~D. Soltani, M.~A. Arfaoui, I.~Tavakkolnia, A.~Ghrayeb, M.~Safari, C.~M.
  Assi, M.~O. Hasna, and H.~Haas, ``{Bidirectional optical spatial modulation
  for mobile users: Toward a practical design for LiFi systems},'' \emph{IEEE
  J. Sel. Areas Commun.}, vol.~37, no.~9, pp. 2069--2086, Sep. 2019.

\bibitem{wang2020constellation}
F.~Wang, F.~Yang, and J.~Song, ``{Constellation optimization under the ergodic
  VLC channel based on generalized spatial modulation},'' \emph{Opt. Exp.},
  vol.~28, no.~14, pp. 21\,202--21\,209, Jul. 2020.

\bibitem{Yesilkaya2019ofdm}
A.~Yesilkaya, R.~Bian, I.~Tavakkolnia, and H.~Haas, ``{OFDM-based optical
  spatial modulation},'' \emph{IEEE J. Sel. Topics Signal Process.}, vol.~13,
  no.~6, pp. 1433--1444, Oct. 2019.

\bibitem{chen2021enhanced}
C.~Chen, X.~Zhong, S.~Fu, X.~Jian, M.~Liu, X.~Deng, and H.~Fu, ``{Enhanced
  OFDM-based optical spatial modulation},'' in \emph{Proc. IEEE IEEE Int. Conf.
  Commun. (ICC)}, Jun. 2021, pp. 1--6.

\bibitem{chen2020ofdm}
\BIBentryALTinterwordspacing
C.~Chen, X.~Zhong, S.~Fu, X.~Jian, M.~Liu, H.~Yang, A.~Alphones, and H.~Y. Fu,
  ``{OFDM-based generalized optical MIMO},'' 2020. [Online]. Available:
  \url{https://doi.org/10.36227/techrxiv.13270751.v1}
\BIBentrySTDinterwordspacing

\bibitem{mesleh2014quadrature}
R.~Mesleh, S.~S. Ikki, and H.~M. Aggoune, ``Quadrature spatial modulation,''
  \emph{IEEE Trans. Veh. Technol.}, vol.~64, no.~6, pp. 2738--2742, Jun. 2015.

\bibitem{komine2004fundamental}
T.~Komine and M.~Nakagawa, ``{Fundamental analysis for visible-light
  communication system using LED lights},'' \emph{IEEE Trans. Consum.
  Electron.}, vol.~50, no.~1, pp. 100--107, Feb. 2004.

\bibitem{chen2020user}
C.~Chen, H.~Yang, P.~Du, W.-D. Zhong, A.~Alphones, Y.~Yang, and X.~Deng,
  ``{User-centric MIMO techniques for indoor visible light communication
  systems},'' \emph{IEEE Syst. J.}, vol.~14, no.~3, pp. 3202--3213, Sep. 2020.

\end{thebibliography}
    
\end{document}